\documentclass[11pt,twoside]{article}
\usepackage{asp2010}
\usepackage{amsfonts,verbatim,graphicx,epsfig,rotating,appendix,srcltx}

\def\fm3{\rm ~fm^{-3}}
	
\newcommand{\beq}{\begin{equation}}
\newcommand{\eeq}{\end{equation}}
\newcommand{\beqa}{\begin{eqnarray}}
\newcommand{\eeqa}{\end{eqnarray}}

\resetcounters
\begin{document}

%opening
\title{Symmetry energy effects in the neutron star properties}
\author{D.E. Alvarez-Castillo, S. Kubis}
\affil{H.Niewodnicza\'nski Institute of Nuclear Physics, Radzikowskiego 152, 
31-342 Krak\'ow, Poland}

\begin{abstract}

The functional form of the nuclear symmetry energy has only been determined in a very narrow range of densities. 
Uncertainties concern both the low as well as the high density behaviour of this function. In this
work different shapes of the symmetry energy, consistent with the
experimental data, were introduced and their  consequences for the crustal
properties of neutron stars are presented. The resulting models are in agreement with astrophysical observations.
\end{abstract}

\section{introduction}
The energy per particle used to described neutron star interios can be expressed in terms of baryon
number  density $n=n_p+n_n$ and isospin asymmetry $\alpha = \frac{n_n-n_p}{n}$
of the system:
\beq
E(n,\alpha)=V(n) + E_s(n)\, \alpha^2 + {\cal O}(\alpha^4)
\label{Enuc}
\eeq
If instead of $\alpha$ the proton fraction $x$ is introduced then then $\alpha =
(1-2x)$, which proves to be useful. In the model used here the only constituents of stellar matter are nucleons and
leptons: electrons and muons. Around and above the nuclear density $n_0 = 0.16 \fm3$
nucleons and leptons form a quantum liquid, which stands for liquid core of the
neutron star. Slightly below $n_0$ matter cannot exist as a homogeneous fluid -
the one-phase system is unstable and the coexistence of two phases is required.
At these densities matter clusterizes  into positive nuclei immersed in a
quasi-free gas of neutrons and electrons, most likely forming a Coulomb lattice with solid state properties and corresponds to the
crust covering the liquid core of a star.

\begin{figure}[htpb!]
\begin{center}$
\begin{array}{cc}
\includegraphics[width=6.5cm]{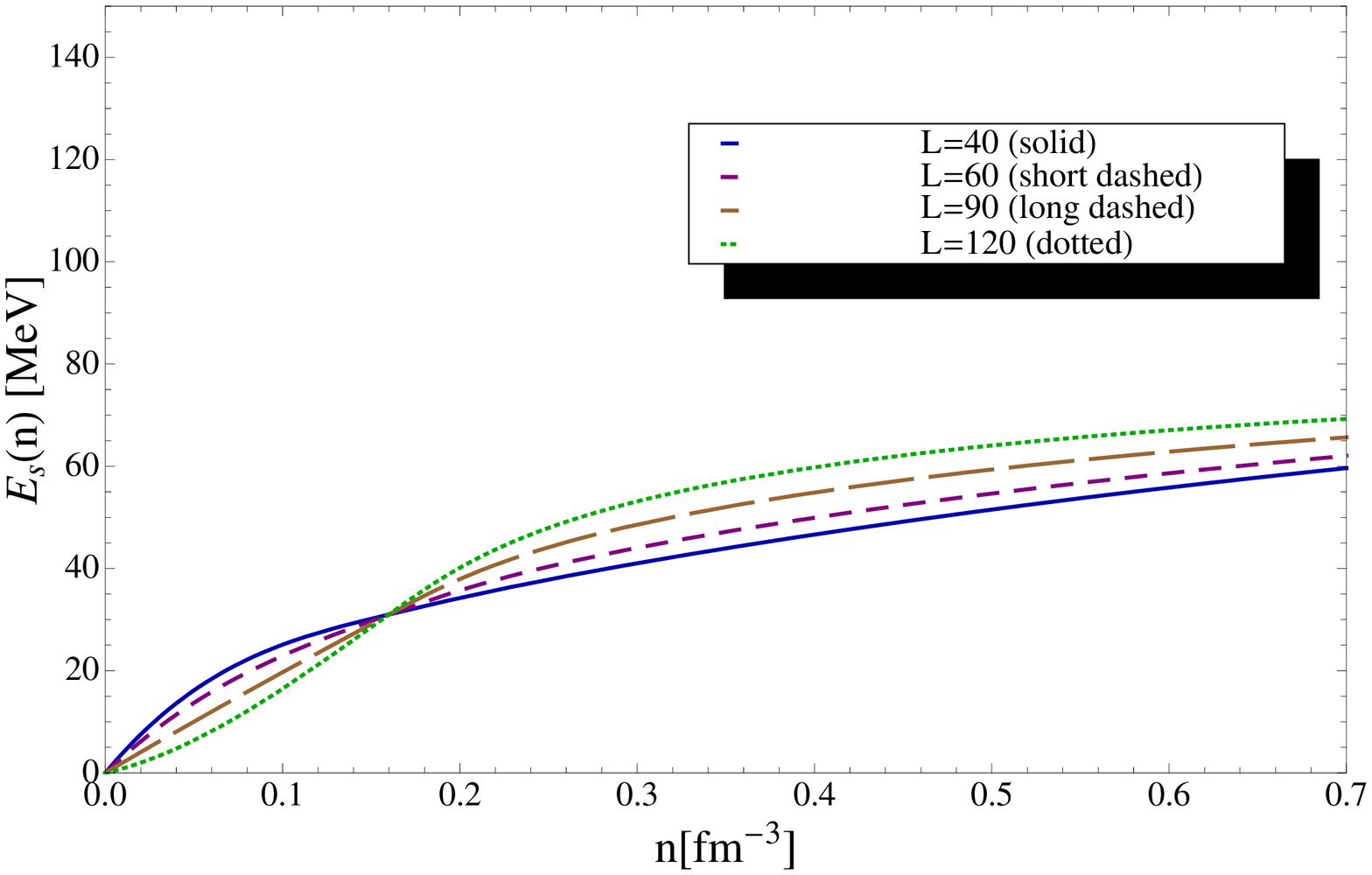} &
\includegraphics[width=6.5cm]{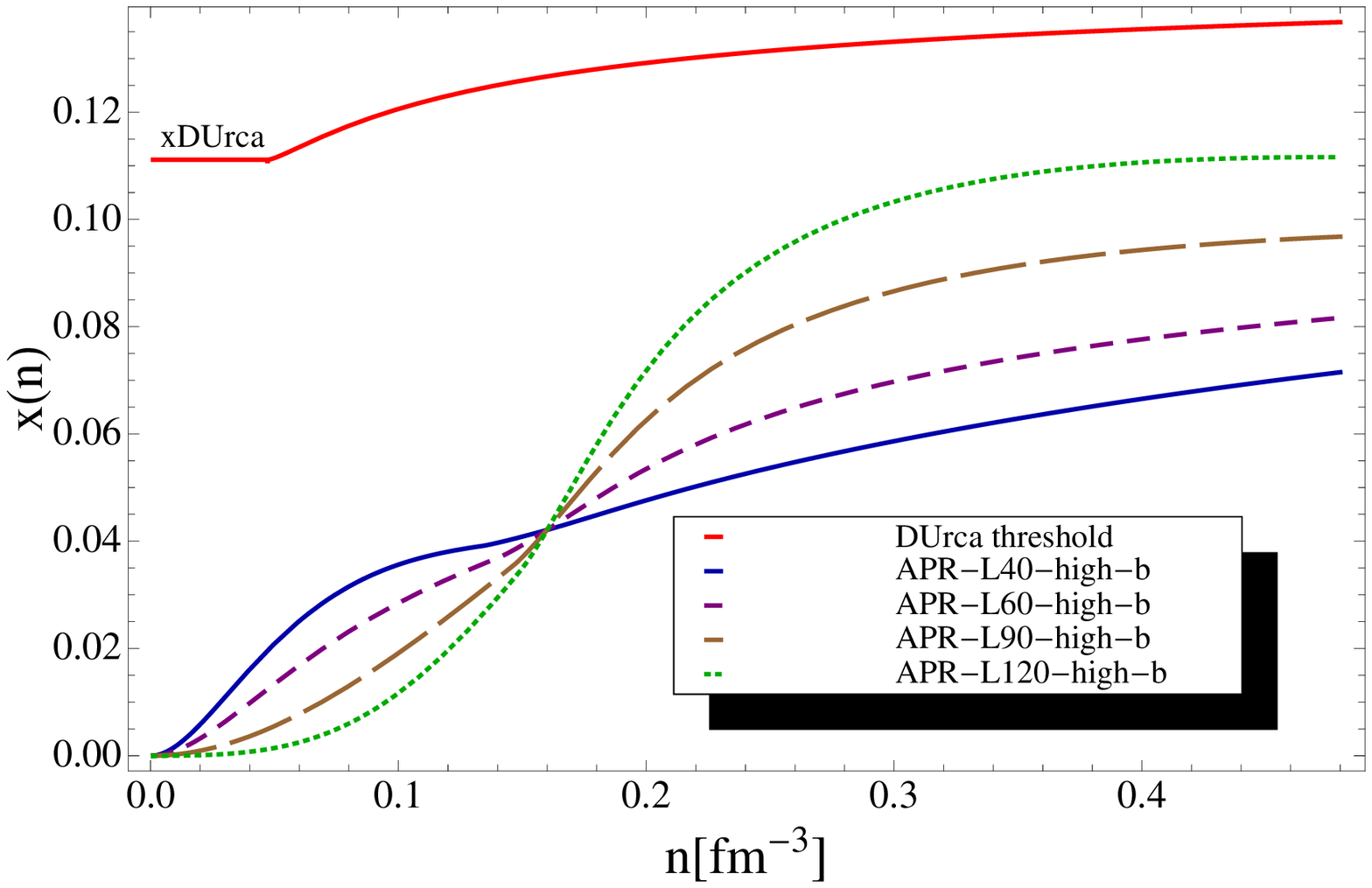}
\end{array}$
\end{center} 
\caption{\textit{Left}. 
Different symmetry energy shapes for the APR-L-high-b models which avoid DUrca
cooling for low NS masses. \textit{Right}. Proton fraction as a function of
baryon number density for these models and DUrca threshold.} 
\label{xDU-L}
\end{figure}

For typical NS masses, between 1-2 M$_\odot$, most of the stellar
matter is occupied by the core, so the global parameters like the mass, radius,
moment of inertia are completely determined by the functional form of  the
Eq.~(\ref{Enuc}). Whereas the isoscalar part $V(n)$ corresponds mainly for the
stiffness of Equation of State (EOS) which is relevant for the maximum mass of
NS,  the isovector part $E_s(n)$ is responsible for the chemical composition of
the matter (see~\citep{2012arXiv1205.6368K} for details).  Both functions $V(n)$
and $E_s(n)$ have been implemented by means of B\'ezier functions composed of
control points ~\citep{Wikipedia:2009:Misc}
\begin{equation}
 \textbf{B}(t)=\sum^{n}_{i=0}{n \choose i}(1-t)^{n-i}t^{i}\textbf{P}_i, \qquad  t \in [0,1]
\end{equation}
where ${n \choose i}$ is the binomial coefficient and the $n+1$ control points
$\textbf{P}_i$, ($i=0,1,2\ldots n$) define the B\'ezier curve of degree $n$. In
this way the isoscalar part $V^{APR}$ follows the shape of the stiffest APR
model (A18+UIX) ~\citep{1998PhRvC..58.1804A} up to 10 times saturation density
$n_0 = 0.16 \fm3$. By use of a B\'ezier curve the model was corrected to fullfill
the saturation point properties, like binding energy -16 MeV and compressibility
$K_0 = 240 $~MeV which were not satisfied by the original A18+UIX. With it, the
most massive neutron star observed of about 2 M$_{\odot}$ 
~\citep{2010Natur.467.1081D} can be created within
this model. For the isovector part, the symmetry energy $E_s$, four different
models sharing the same high density behaviour but having different slopes at
saturation density have been introduced. The measured values of $E_s(n_0)$ and
$L$ (related to the slope of $E_s$) of about 30 MeV and 40-120 MeV are used
for drawing the $E_s$ curves. All of them satisfy the DUrca constraint that
dictates that low mass neutron stars must not cool by DUrca
process~\citep{2006A&A...448..327P} related to the proton fraction $x$ inside
the star, which must stay below the DUrca proton fraction threshold $x$.
Figure~\ref{xDU-L} shows these symmetry energy forms and the resulting proton
fractions together with the DUrca threshold. Control points for $V^{APR}$ are
presented in table~\ref{V-APR} whereas for $E_s$ can be found
in~\citep{2012arXiv1205.6368K}.

\section{Neutron star properties}

To determine the neutron star properties two EoS describing its core and its crust are joint at equal pressure and density. For the crust the 
SLy EOS composed of different parts whose table can be  found in~\citep{Ioffe:2012:Misc} has been used. The liquid core is described by Eq.~(\ref{Enuc}) and the B\'ezier
curves described above. To determine the crust-core transition three different methods are used which are presented in~\citep{2012arXiv1205.6368K} and the resulting transition
densities $n_c$ are presented in table~\ref{nccLmodels}.
\begin{table}[ht!]
\center
\caption{Crust-core transition densities for the {\it APR-L-high-b} models.}
\label{nccLmodels}
\begin{tabular}{cccc}
 \hline \hline
  model & $n_{c}(Q)$ & $n_{c}(K_{\mu})$ & $n_{c}(1\leftrightarrow 2)$ \\
\hline
 APR-L40-high-b &  0.103076 & 0.11012  		& 0.116185 \\
 APR-L60-high-b & 0.0922071 &	0.101941	& 0.104926 \\
 APR-L90-high-b & 0.0870974 &	0.102523	& 0.103922 \\
 APR-L120-high-b& 0.115633  &	0.142939	& 0.147017 \\
\hline \hline
\end{tabular}
\end{table}
\begin{figure*}[thb!]
\includegraphics[width=7cm]{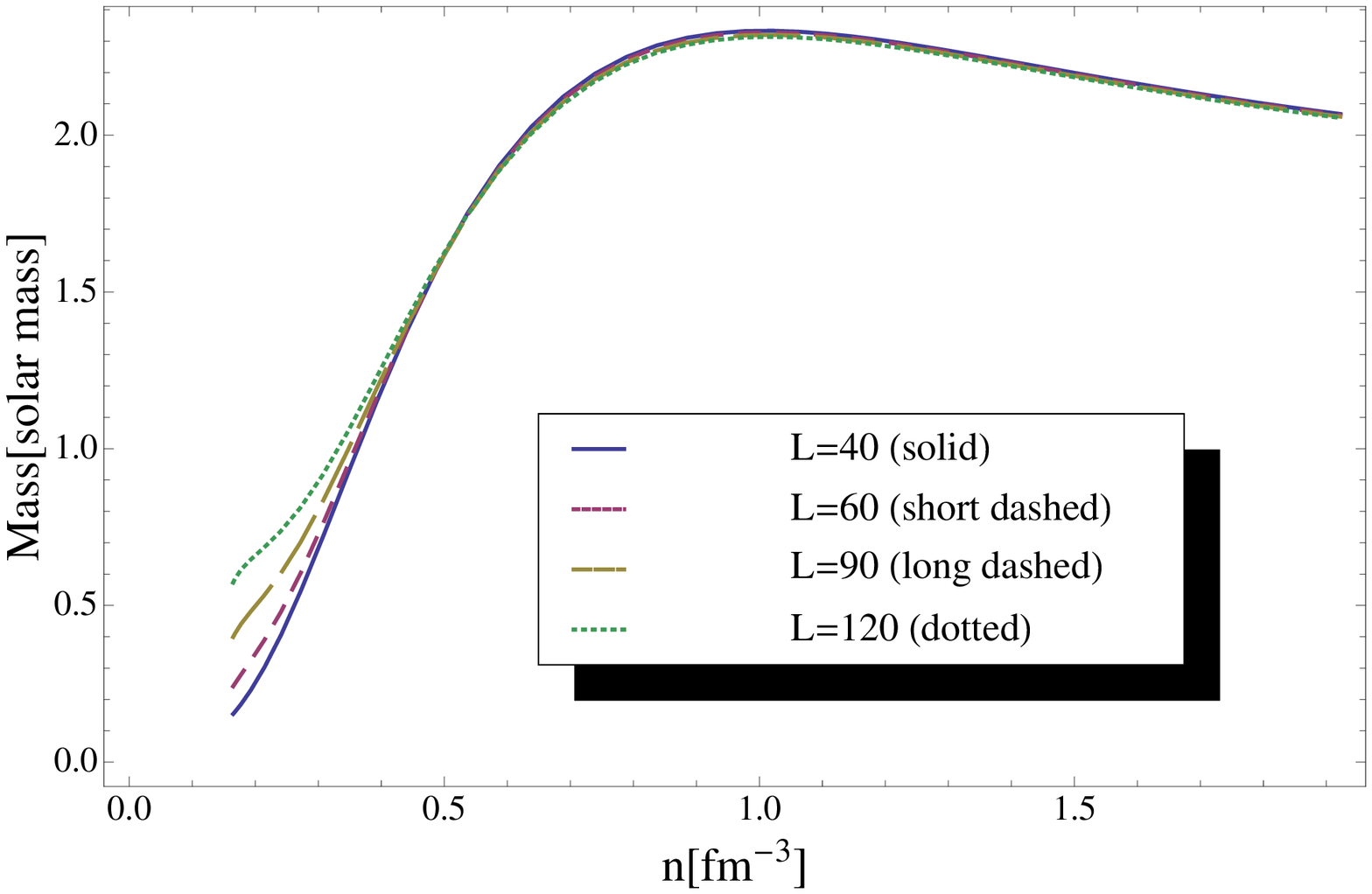}
\includegraphics[width=7cm]{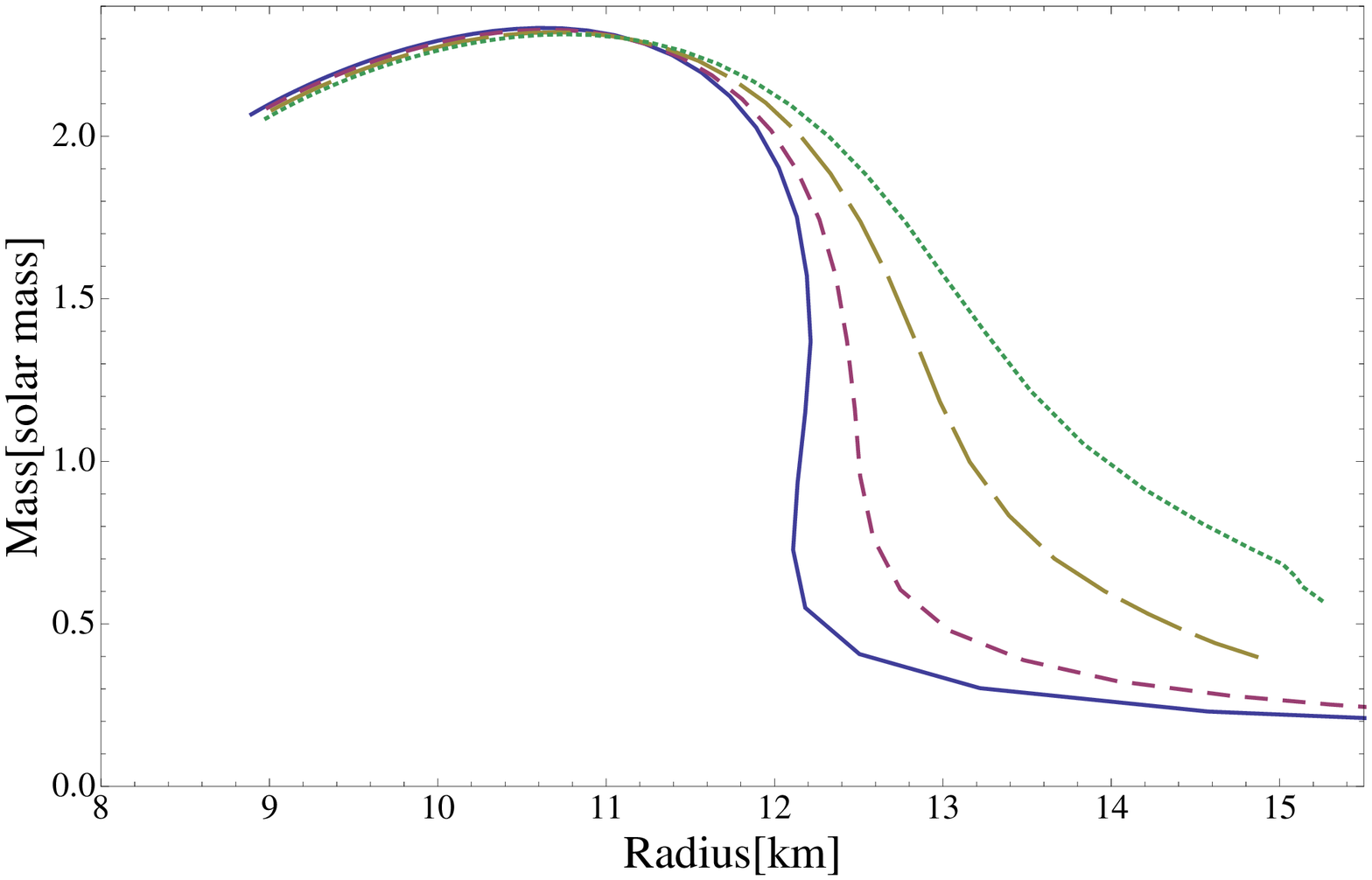}

\includegraphics[width=7cm]{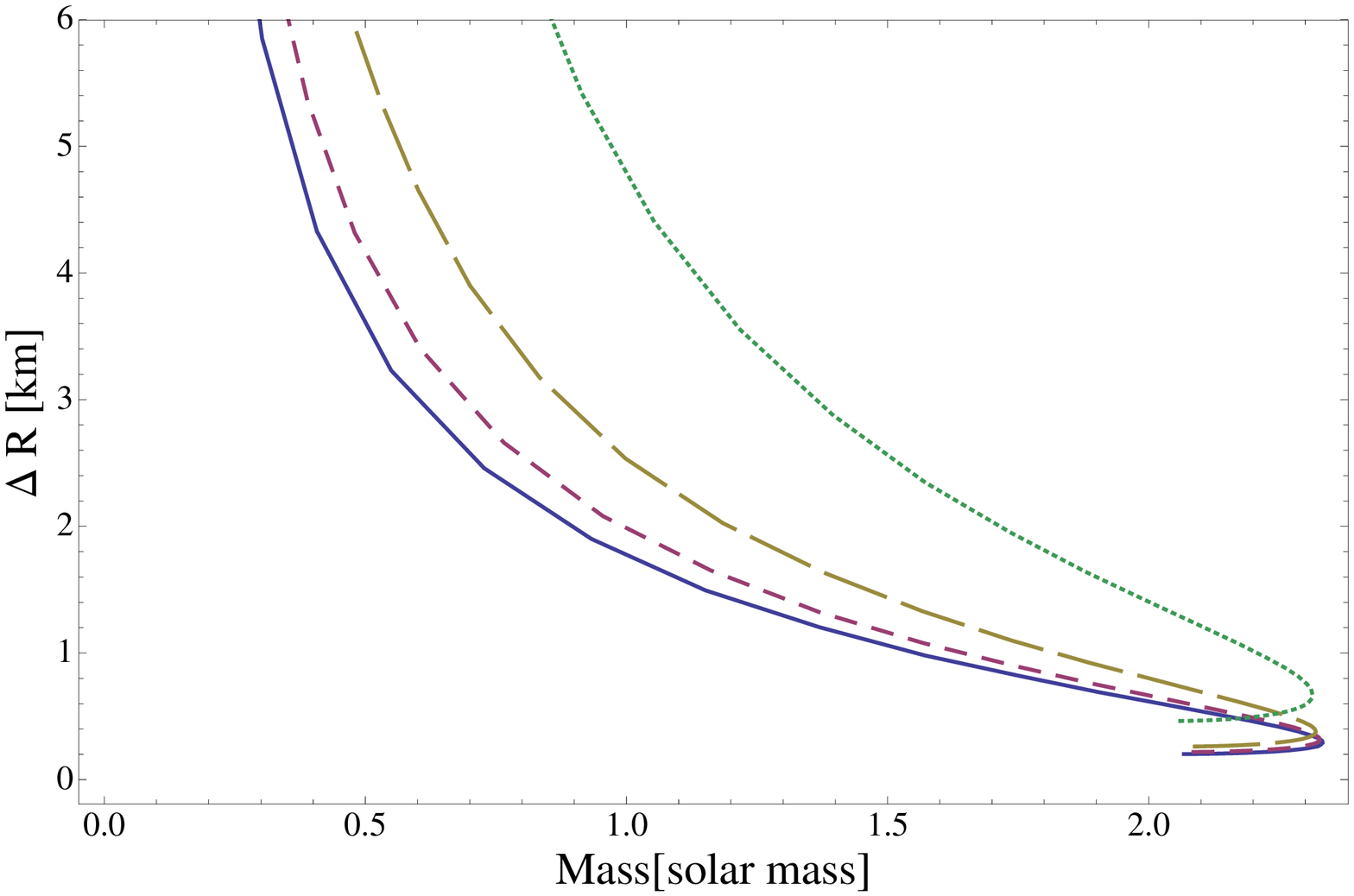}
\includegraphics[width=7cm]{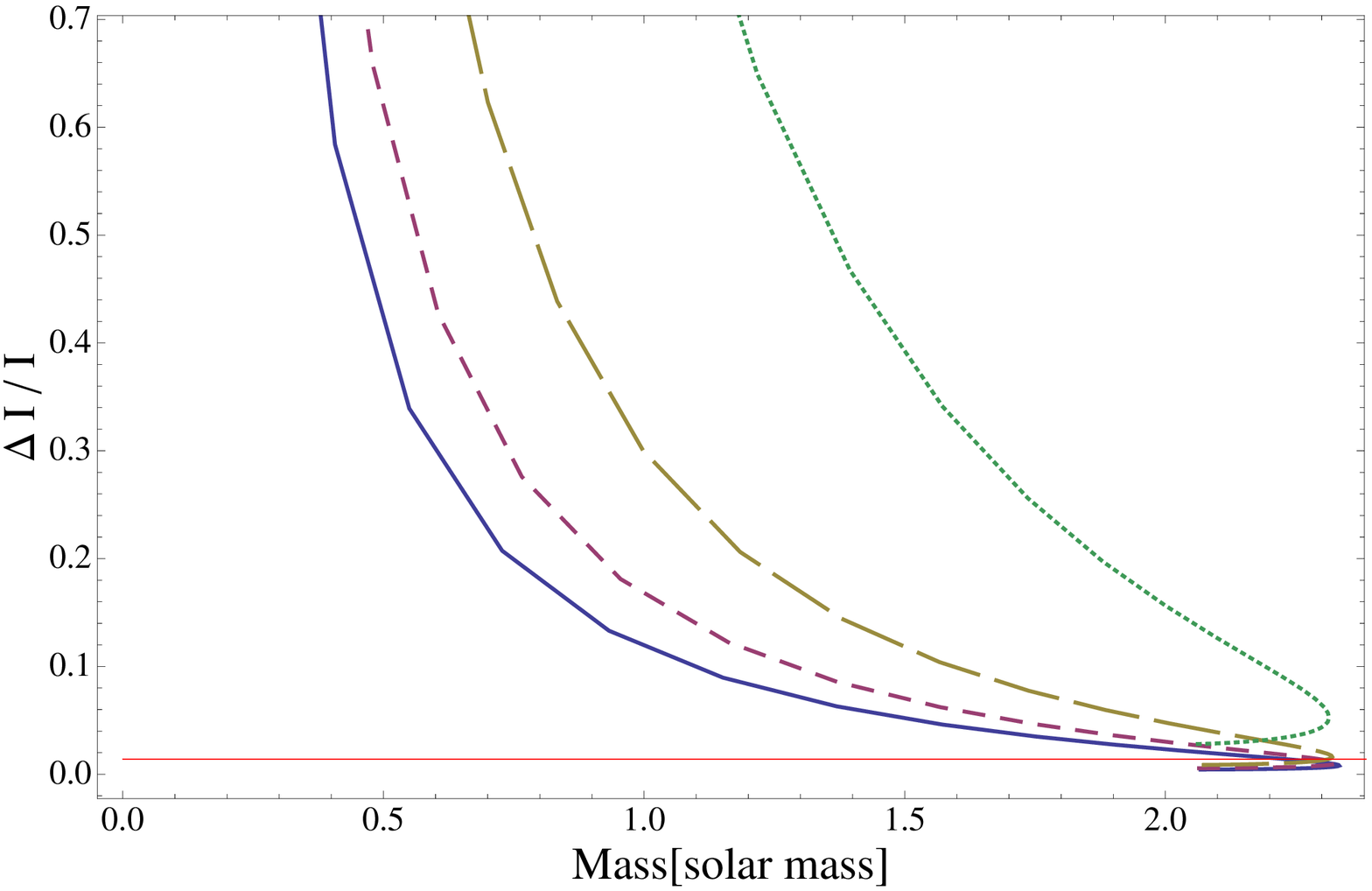}
\caption{Neutron star features for the  APR-L-high-b models. The horizontal line in the right lower figure represents the glitch constraint from~\citep{1999PhRvL..83.3362L}.}
\label{Lvar}
\end{figure*}
The macroscopic properties like radius, mass, moment of inertia are derived in the framework of General Relativity
by solving the TOV equations and the crust thickness is derived by the use of $n_c(K_{\mu})$ as described in~\citep{2011AIPC.1396..165A}.
Figure 2 shows the resulting neutron star properties for each family. They all produce high enough masses and present thick crusts. In particular
the moment of inertia carried by the crust does not impose stringent constraints due to the glitch model restriction~\citep{1999PhRvL..83.3362L}.

It is important to mention that the inclusion of $V^{APR}$ in these models is crucial for creating massive enough neutron stars. Furthermore the particular 
form of $E_s$ fulfills the DUrca requirement. These both constraints cannot be satisfied in paralell with other isoscalar parts not so stiff, like is the
case of the PAL parametrization presented in~\citep{2012arXiv1205.6368K}. Therefore the models here are good candidates for the EoS that can be compared
to microscopic approaches. Lately a new study that combined different 
laboratory measurements points out
values  of $E_s(n_0)
\approx 32 ~\textrm{MeV},~~L\approx 50 ~\textrm{MeV}$ with an error of a
few MeV~~\citep{2012arXiv1203.4286L}. From that, one may conclude that models with low $L$ values are preferable.

\begin{table}[ht!]
\label{V-APR}
\caption{
B\'ezier control points for the $V^{APR}$ isoscalar function which follows the
APR A18+UIX EoS at high $n$. }
\center
\begin{tabular}{ccccc}
 \hline \hline
   $\textbf{P}_0$ & $\textbf{P}_1$ & $\textbf{P}_2$ & $\textbf{P}_3$ & \\
\hline
  (0.0016, 0.4649) & (0.08, -20.9676) &  (0.16, -14.6553) & (0.24, -27.2516)\\ 
\hline \hline 
 $\textbf{P}_4$ & $\textbf{P}_5$ & $\textbf{P}_6$& $\textbf{P}_7$ & \\
\hline
 (0.48, -29.7266) & (0.8, 86.6426)& (1.12, 320.0992) & (1.6, 1093.7393) \\
\hline \hline
\end{tabular}
\end{table}

\acknowledgements This work has been partially supported by CompStar a research networking
programme of the European Science Foundation. D.E. Alvarez-Castillo thanks the organizers for hospitality and attention.

\bibliography{myrefs}

\end{document}